\documentclass[prb,twocolumn,showpacs,superscriptaddress]{revtex4}

\usepackage{graphicx}
\usepackage{amsmath}
\usepackage{amssymb}
\usepackage{epsfig}

\usepackage{ctable}
\usepackage{slashbox}

\renewcommand{\v}[1]{{\bf #1}}

\def\eqa{\begin{eqnarray}}
\def\eea{\end{eqnarray}}
\newcommand{\eq}{\begin{equation}}
\newcommand{\ee}{\end{equation}}
\newcommand{\nn}{\nonumber\\}

\newcommand{\<}{\langle}
\renewcommand{\>}{\rangle}

\newcommand{\ua}{\uparrow}
\newcommand{\da}{\downarrow}
\newcommand{\ra}{\rightarrow}

\newcommand{\al}{\alpha}
\newcommand{\bt}{\beta}
\newcommand{\del}{\delta}
\newcommand{\Del}{\Delta}

\newcommand{\ga}{\gamma}
\newcommand{\Ga}{\Gamma}

\newcommand{\la}{\lambda}
\newcommand{\La}{\Lambda}

\newcommand{\si}{\sigma}

\usepackage{color}

\begin{document}

\title{Possible Superconductivity in Electron-doped Chromium Pnictide LaOCrAs}

\author{Wan-Sheng Wang}
\affiliation{Department of Physics, Ningbo University, Ningbo 315211, China}
\email{wangwansheng@nbu.edu.cn}

\author{Miao Gao}
\affiliation{Department of Physics, Ningbo University, Ningbo 315211, China}

\author{Yang Yang}
\affiliation{College of Physics and Electronic Engineering,
Zhengzhou University of Light Industry, Zhengzhou 450002, China}

\author{Yuan-Yuan Xiang}
\affiliation{College of Science, Hohai University, Nanjing, 210098, China}

\author{Qiang-Hua Wang}
\affiliation{National Laboratory of Solid State Microstructures and School of Physics,
Nanjing University, Nanjing, 210093, China}
\affiliation{Collaborative Innovation Center of Advanced Microstructures,
Nanjing University, Nanjing, 210093, China}
\email{qhwang@nju.edu.cn}

\begin{abstract}
We constructed an effective tight-binding model with five Cr $3d$ orbitals for 
LaOCrAs according to first-principles calculations. Basing on this model, we investigated
possible superconductivity induced by correlations in doped LaOCrAs using the functional
renormalization group (FRG). We find that there are two domes of superconductivity in electron-doped LaOCrAs. With
increasing electron doping, the ground state of the system evolves from G-type antiferromagnetism in the parent
compound to an incipient $s_\pm$-wave superconducting phase dominated by electron bands derived from the $d_{3z^2-r^2}$ orbital as the filling is above $4.2$ electrons per site on the $d$-orbitals of Cr. The gap function has strong octet anisotropy on the Fermi pocket around the zone center and diminishes on the other pockets. In electron over-doped LaOCrAs, the system develops $d_{x^2-y^2}$-wave superconducting phase and the active band derives from the $d_{xy}$ orbital. Inbetween the two superconducting domes, a time-reversal symmetry breaking $s+id$ SC phase is likely to occur. We also find $s_\pm$-wave superconducting phase in the hole-doped case.
\end{abstract}




\maketitle

\section{Introduction}

Transition metal pnictides have received
much attention since the discovery of high-$T_c$ superconductivity
in F-doped LaOFeAs. \cite{Hosono1} Along with the extensively studied iron pnictides, some iron-free isostructural compounds have been found to be superconductors, such as pnictides based on Ni, Rh, Ir, Pt, and Pd, etc.
\cite{Ni1,Ni2,Ni3,Ni4,Ni5,Ni6,Rh1,Rh2,Ir,Pd1,Pd2,Pt1,Pt2}
However, Mn-based pnictides are antiferromagnetic (AFM) insulators.\cite{Mn1,Mn2,Mn3,Mn4,Mn5,Mn6,Mn7}
No superconductivity has been found when the antiferromagnetism is
suppressed by pressure in LaOMnP \cite{LaOMnP} or by high electron doping through H$^-$
substitution of O$^{2-}$ in LaOMnAs. \cite{LaOHMnAs} It is likely because the strong Hund's rule coupling
in the half-filled $d^5$ configuration forces the Mn-ion to form a high-spin state that is always toxic to superconductivity. With the $d^5$ configuration as the reference, it would be interesting to look for superconductivity in compounds with $d^4$ configuration, such as in the isostructural chromium pnictides, which mirrors the $d^6$ configuration in parent iron pnictides.\cite{LaOCrAs,BaCr2As2_3} This motivation makes better sense after the observation of superconductivity in CrAs (under high pressures) \cite{CrAs1,CrAs2} and $A_2$Cr$_3$As$_3$ ( $A =$ K, Rb, Cs) under ambient pressure.\cite{ACrAs1,ACrAs2,ACrAs3}

In fact, BaCr$_2$As$_2$ and SrCr$_2$As$_2$ were synthesized in 1980, \cite{BaCr2As2_1} and the physical
properties and electronic structure of BaCr$_2$As$_2$ were investigated \cite{BaCr2As2_2}
after the discovery of superconductivity in doped BaFe$_2$As$_2$. \cite{BaFe2As2}
The results show that BaCr$_2$As$_2$ is an AFM metal,
with a G-type order and very strong magnetic interactions.
In addition, an analogous compound EuCr$_2$As$_2$ with similar
electronic properties were reported. \cite{EuCr2As2} A series of
$Ln$OCrAs ($Ln =$ La, Ce, Pr, and Nd) compounds were synthesized by Park \emph{et al}. \cite{LnOCrAs}
All the members show metallic electronic conduction, and LaOCrAs is ordered
in G-type antiferromagnetism with a large spin moment of 1.57$\mu_B$ along the $c$ axis
revealed by powder neutron diffraction under 300K. Recently, a new chromium
oxypnictide Sr$_2$Cr$_3$As$_2$O$_2$ containing CrO$_2$ and Cr$_2$As$_2$
square-planar lattices were reported. It is also an AFM metal with a G-type
AFM order in the Cr$_2$As$_2$ plane under 291K.\cite{SrCrAsO} However, superconductivity
has not yet been observed in any of the above parent Cr$_2$As$_2$-layer based materials.
But as in the case of iron pnictides, superconductivity may appear upon doping the parent
compounds to a sufficient level.

In this work, we investigate possible superconductivity in doped chromium
pnictide LaOCrAs. First, we calculate the electronic structure of LaOCrAs
by first-principles calculations and construct an effective five-orbital
tight-binding model. Second, on the basis of the tight-binding model, we
investigate possible superconductivity induced by correlations in
doped LaOCrAs using the unbiased singular-mode functional renormalization
group (SMFRG). \cite{wws1,xyy1,xyy2,wws2,xyy3,yy,sro,yy2,wws3}
In the parent compound, we confirm the G-type antiferromagnetism found in experiments.
Upon electron doping, We find that there are two domes of superconductivity as the
filling is above $4.2$ electrons per site (henceforth on the $d$-orbitals of Cr).
The first superconducting phase has an incipient $s_\pm$-wave pairing symmetry, with gap sign change on the electron Fermi pockets and the virtual hole pockets. The gap function has octet anisotropy on the Fermi pocket around the zone center, and diminishes on the other pockets. This superconducting phase
is triggered by spin fluctuations between the Fermi pockets around
$\Ga$ and $M$ points in the unfolded Brillouin Zone (BZ),
where the orbital contents of the Bloch states are dominant by $d_{3z^2-r^2}$.
With further electron doping, the system
develops $d_{x^2-y^2}$-wave superconducting phase, which
is related to spin fluctuations between the Fermi pockets around $X$
and $Y$ points, where the orbital characters are
dominant by $d_{xy}$. Inbetween the two superconducting phases, a time-reversal-symmetry breaking
$s+id$ phase would be energetically favorable. In the hole doped case we find
$s_\pm$-wave superconductivity is favorable.

The rest of the paper is organised as follows. In Sec.\ref{mm}, we
construct an effective model for LaOCrAs through first-principles
calculations and briefly introduce the SMFRG method. In Sec.\ref{rd}, we discuss the
results for doped LaOCrAs and conclude by a phase diagram. Finally, a
summary and experimental perspectives are discussed in Sec.\ref{summary}.

\section{model and method}\label{mm}

LaOCrAs has a ZrCuSiAs-type structure with the space group $P4/nmm$ (No. 129).
Because of the tetrahedral coordination of As, there are two Cr atoms per unit
cell. The experimentally determined lattice constants are $a=4.04123$\AA$ $ and $c=8.98637$\AA . We obtain the band structure by the Quantum-ESPRESSO
package,\cite{LDA} and then construct the maximally localized Wannier functions.\cite{Wannier}
These maximally localized Wannier functions are centered at two Cr sites
in the unit cell, transforming as $d_{3Z^2-R^2}$, $d_{XZ}$,
$d_{YZ}$, $d_{XY}$, and $d_{X^2-Y^2}$, where $X$, $Y$, $Z$ refer to the axes of the large
unit cell. Since the two Cr atoms satisfy the group symmetry,
we further unfold the BZ with one Cr atom per small unit cell and construct a
five-orbital model. For convenience, we rotate the crystal coordinates axes and the
orbitals basis by $\pi/4$, form $X$-$Y$ to $x$-$y$.\cite{Kuroki} Since
the inter-layer coupling is weak, we only consider the in-plain hopping
for brevity. The in-plain hopping integrals $t_{\Del_x,\Del_y}^{\mu,\nu}$ are displayed
in Table.\ref{hoppings}, where $[\Del_x,\Del_y]$ denote the in-plain hopping
vector, and $\mu$, $\nu$ label the five rotated $d$-orbitals: $d_{3z^2-r^2}$,
$d_{xz}$, $d_{yz}$, $d_{x^2-y^2}$, and $d_{xy}$. They are directed along the nearest Cr-Cr bonds.

\begin{table*}
\caption{ Hopping integrals $t_{\Del_x,\Del_y}^{\mu,\nu}$ (in units of $eV$)
for effective five Cr $3d$ bands model. $[\Del_x,\Del_y]$ denotes the in-plain
hopping vector, and $(\mu,\nu)$ the orbitals. $\sigma_y$, $I$ and $\sigma_d$
are three basic group operation, corresponding to t$_{-\Del_x, \Del_y}^{\mu,\nu}$,
t$_{-\Del_x,-\Del_y}^{\mu,\nu}$, and t$_{\Del_y,\Del_x}^{\mu,\nu}$, respectively.
Notice that the $x$ and $y$ axes are along the nearest Cr-Cr bond. The basis
of the orbitals are also along the nearest Cr-Cr bond. The order of the
five orbitals are:($d_{3z^2-r^2}$, $d_{xz}$, $d_{yz}$, $d_{x^2-y^2}$, $d_{xy}$),
with the corresponding on-site energies (11.513,11.795,11.795,11.455,11.967)eV,
respectively. The chemical potential is 11.486 eV.}
\centering
\begin{tabular*}{\textwidth}{p{2.5cm} *{5}{p{2cm}<{\centering}} *{3}{p{1.4cm}<{\centering}}}
    \hline \hline
    \backslashbox{$(\mu,\nu)$}{$(\Del_x, \Del_y)$} & (1, 0) & (1, 1) & (2, 0) & (2, 1) & (2, 2)
    & $\sigma_{y}$ & $I$ & $\sigma_{d}$ \\ \hline
    $(1,1)$ & -0.012 &  0.037 & -0.048 & -0.001 & -0.009 & + & + & + \\
    $(1,2)$ & -0.095 &  0.076 &  0.012 &  0.013 &  0.008 & - & - & (1,3) \\
    $(1,3)$	&        &  0.076 &        &  0.022 &  0.008 & + & - & (1,2) \\
    $(1,4)$	& -0.263 &        & -0.038 & -0.014 &        & + & + & - \\
    $(1,5)$	&        & -0.104 &        & -0.005 &  0.003 & - & + & + \\
    $(2,2)$ & -0.300 &  0.183 & -0.018 &  0.013 &  0.014 & + & + & (3,3) \\
    $(2,3)$ & 	     &  0.047 & 	   &  0.016 &  0.006 & - & + & + \\
    $(2,4)$ & -0.299 &  0.078 & -0.017 & -0.012 & -0.001 & - & - & -(3,4) \\
    $(2,5)$ &        &  0.055 &        &  0.010 & -0.003 & + & - & (3,5) \\
    $(3,3)$ &  0.122 &  0.183 &  0.002 &  0.003 &  0.014 & + & + & (2,2) \\
    $(3,4)$ &        & -0.078 & 	   & -0.012 &  0.001 & + & - & -(2,4) \\
    $(3,5)$ &  0.257 &  0.055 &  0.023 &  0.019 & -0.003 & - & - & (2,5) \\
    $(4,4)$	&  0.415 & -0.025 & -0.007 & -0.004 &  0.010 & + & + & + \\
    $(4,5)$ &        &        &        & -0.017 &        & - & + & - \\
    $(5,5)$ &  0.036 &  0.100 & -0.012 & -0.030 & -0.030 & + & + & + \\
  \hline
\end{tabular*}
\label{hoppings}
\end{table*}

\begin{figure}
\includegraphics[width=9cm]{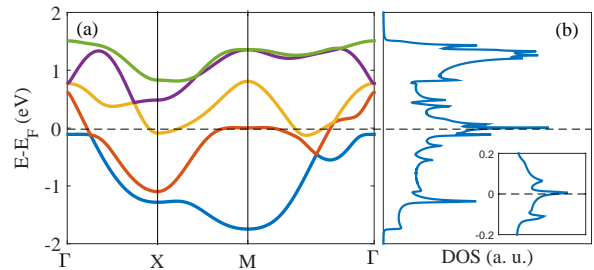}
\caption{(Color online) (a) Band structure of LaOCrAs based on the
five-band model in the unfolded Brillouin zone. (b) The corresponding density of states.
Inset shows the enlarged window near Fermi level. }\label{dis}
\end{figure}

\begin{figure}
\includegraphics[width=8cm]{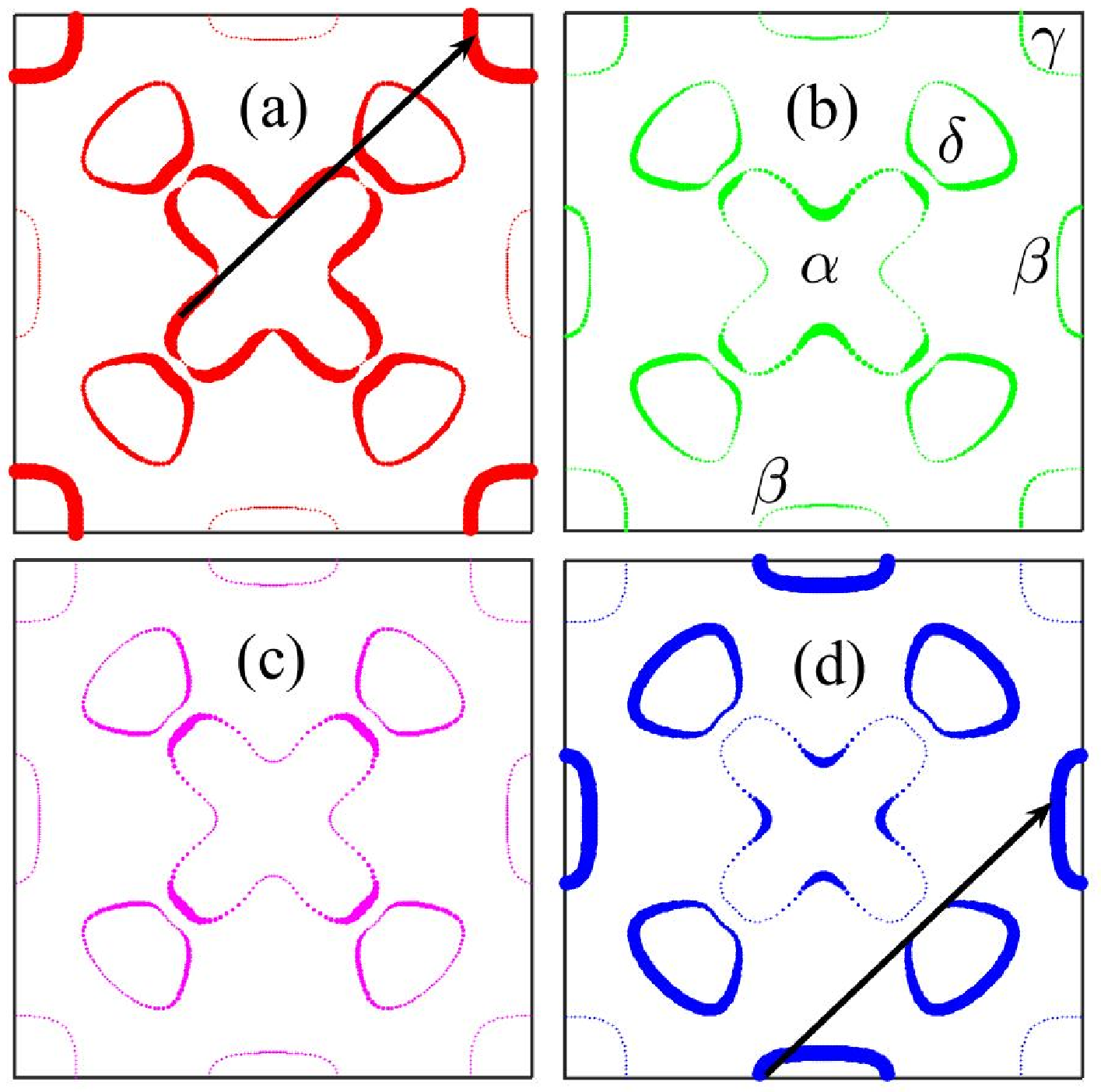}
\caption{(Color online) Fermi surface (FS) of LaOCrAs in the unfolded Brillouin zone.
From (a)-(d) the width of each FS line is proportional to its spectral weights in
$d_{3z^2-r^2}$, $d_{xz}$, $d_{x^2-y^2}$, and $d_{xy}$ components.
The arrows in (a) and (d) denote two types of
scattering with momentum $\v Q$ around $(\pi, \pi)$. The $\al$, $\bt$, and $\ga$
in (b) denote the Fermi pockets around $\Ga$, $X$(or $Y$), and $M$ points, respectively.
The pockets around the middle points of $\Ga$-$M$ lines are denoted by $\del$. }\label{fs}
\end{figure}

We obtain the band structure in the unfolded BZ as shown in Fig.\ref{dis}(a).
The bands close to Fermi level around $\Ga$ and $M$
points are nearly flat, resulting in large density of states (DOS) near
Fermi level, as shown in the Fig.\ref{dis}(b), where the DOS has two
peaks around the Fermi level. The large DOS near Fermi
level makes the system particularly susceptible to various instabilities driven by
electron correlations, as will be discussed later. Slightly above the two peaks, there is also
a peak in DOS due to the van Hove singularities in the third band.
Figs.\ref{fs}(a)-(d) show the Fermi surface (FS) of LaOCrAs encoded with the spectral weights in $d_{3z^2-r^2}$, $d_{xz}$, $d_{x^2-y^2}$, and $d_{xy}$ components. The $d_{yz}$ weight is not shown since
it is equivalent to that of $d_{xz}$ upon a rotation of $\pi/2$.
We see that the $\al$ and $\ga$ pockets are dominant by
$d_{3z^2-r^2}$ orbital, and in particular the weight has octet-wise deep minima on $\al$. The $\bt$ pockets are dominant by $d_{xy}$ orbital, and finally
the $\del$ pockets by $d_{3z^2-r^2}$ and $d_{xy}$ orbitals.


We consider the following local interactions on Cr atoms:
\eqa
 H_{I} && = U \sum_{i\mu} n_{i\mu\ua}n_{i\mu\da}
     +U' \sum_{i,\mu>\nu} n_{i,\mu}n_{i,\nu} \nn
    && +J_{H}\sum_{i,\mu>\nu,\si\si'}c^{\dag}_{i\mu\sigma}c_{i\nu\sigma}
        c^{\dag}_{i\nu\sigma'}c_{i\mu\sigma'} \nn
    && +J'_{H}\sum_{i,\mu\neq \nu}c^{\dag}_{i\mu\ua}c^{\dag}_{i\mu\da}
    c_{i\nu\da}c_{i\nu\ua},
\label{H}
\eea
where $i$ denotes Cr sites, $\si$ is the spin polarity, $\mu$ and $\nu$
denote five Cr $3d$ orbitals, $n_{i\mu}=\sum_{\mu\si} c_{i\mu\si}^\dag
c_{i\mu\si}$, $U$ is the intra-orbital repulsion, $U'$ is the inter-orbital
repulsion, $J_H$ is Hund's rule coupling, and $J'_H$ is the pair hopping term,
and we use the Kanamori relations $U'=U-2J_H$ and $J_H = J'_H$ so that we are
left with two independent interaction parameters
$(U, J_H)$. The interactions can lead to competing collective fluctuations
in density-wave and pairing channels, which we handle by SMFRG as follows.
A general interaction vertex function can be decomposed as scattering matrices
between composite bosons,
\eq
V^{\mu, \nu; \la , \eta}_{\v k, \v k', \v q} \ra \sum_m S_m(\v q)
\phi^{\mu, \nu}_m(\v k, \v q) [\phi^{\la, \eta}_m(\v k', \v q)]^*,
\ee
either in the superconducting (SC), spin-density wave (SDW), or charge-density
wave (CDW) channels. Here, $(\mu, \nu, \la, \eta)$ are orbital indices,
$\v q$ is the collective momentum, and $\v k$ ( or $\v k'$) is an internal
momentum of the Fermion bilinears $c^\dag_{\v k + \v q, \mu}
c^\dag_{-\v k, \nu}$ and $c^\dag_{\v k + \v q, \mu} c_{\v k ,\nu}$
in the particle-particle and particle-hole channels, respectively.
In the following we define, in a specific channel, $ S(\v q)$ as the
leading attractive eigenvalue at $\v q$, and $S$ as the globally
leading one. The SM-FRG provides the \emph{coupled} flow of all channels
versus a decreasing energy scale $\La$ (the infrared limit of the Matsubara
frequency in our case). The fastest growing eigenvalue $S(\v Q)$ implies
an emerging order associated with a collective wave vector $\v Q$ and
an eigenfunction (or form factor) $\phi(\v k, \v Q)$. (Notice that $\v Q =0$
in the SC channel because of the Copper instability, but may evolve with
$\La$ in the other channels.) The divergence scale provides an upper limit
of the ordering temperature. More technical details can be found elsewhere.
\cite{wws1,xyy1,xyy2,wws2,xyy3,yy,sro,yy2,wws3}

\section{Results and Discussion}\label{rd}

\begin{figure}
\includegraphics[width=9cm]{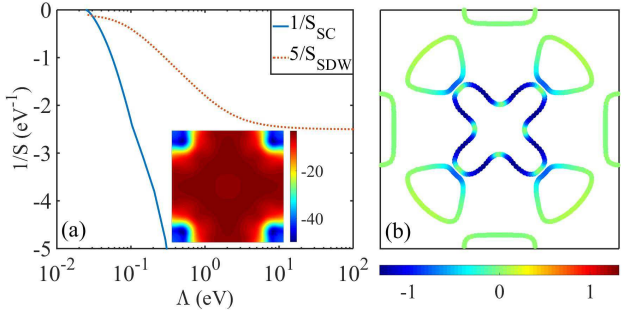}
\caption{(Color online) Results for $n=4.24$ with $U=1.0$eV and $J_H=U/4$.
(a) FRG flow of $1/S_{SC,SDW}$, the inverse of the leading attractive
interactions, versus the running energy scale $\La$. Notice that
$1/S_{SC,SDW} \ra 0^-$ if $S_{SC,SDW}$ diverges. The inset shows
$S_{SDW}(\v q)$ in the unfolded Brillouin zone at the final energy
scale. (b) Fermi surface and gap function $\Del (\v k)$ (color scale). }
\label{ssc}
\end{figure}

We first discuss the electron-doped case at band filling $n=4.24$,
with the Fermi level slightly above the flat band around the $M$ point. The
$\ga$ pocket is absent here as shown in Fig.\ref{ssc}(b),
but we should emphasize that our SMFRG includes virtual excitations from all
bands. Fig.\ref{ssc}(a) shows the FRG flow of the leading eigenvalues
$S_{SC, SDW}$ versus the running energy scale $\La$ for $U=1.0 eV$ and
$J_H = U/4$. Since the CDW channel remains weak during the flow, we shall
not discuss it henceforth. We find that the SDW channel is enhanced in the
intermediate stage, but saturate at low energy scales because of lack
of phase space for low-energy particle-hole excitations. The momentum
$\v q$ associated with the leading $S_{SDW}$ is around $(\pi, \pi)$ at
high energy scales and only changes slightly during the flow. The inset of
Fig.\ref{ssc}(a) shows $S_{SDW}(\v q)$ versus $\v q$ at the final stage.
There are peaks around $\v Q = (\pi, \pi)$. We checked that the associated
form factors describe site-local spins, indicating G-type AFM fluctuations,
consistent with the G-type AFM order in the parent compound. \cite{LnOCrAs}
This G-type spin fluctuations can be associated with two types of scattering
as shown in Fig.\ref{fs}.

The enhancement of $S_{SDW}$ in the intermediate stage triggers attractive
pairing interaction $S_{SC}$ to increase, and the latter diverges eventually on its
own via the Copper mechanism. Thus, the system develops an SC instability
at the divergence energy scale as shown in Fig.\ref{ssc}(a). Since the
gap functions of the singlet SC state would change sign on the two $\v k$ points
connected by the spin fluctuation vector $\v Q$, there are two competing pairing states:
(i) Incipient $s_{\pm}$-wave, with sign change of the gap function on the $\al$ and $\ga$ pockets.
(Note the $\ga$ pocket is slightly bellow the Fermi level here.) (ii) $d_{x^2-y^2}$-wave, with sign
change on the two $\bt$ pockets. By checking the pairing form factor (see the Appendix \ref{form} for details)
we find the incipient $s_{\pm}$-wave is realized at the present doping level. \cite{wanyuan} The gap function
in the band basis (see Appendix \ref{form}) is shown along the Fermi surfaces in Fig.\ref{ssc}(b).  We observe that the
gap function on the $\al$ pocket has octet-wise deep minima, and the amplitude roughly scales with
the $d_{3z^2-r^2}$-weight of the normal state on this pocket. This is also the case on
the $\del$ pockets. More interestingly, the gap function on the $\bt$ pockets diminishes, which is consistent with the frustration caused by the spin scattering between $\bt$ pockets that would favor $d_{x^2-y^2}$-wave pairing instead. Therefore, we obtained a strongly orbital-selective pairing, and the active orbital is $d_{3z^2-r^2}$.
We notice that similar octet-wise gap minima appears in heavily hole doped Ba$_{1-x}$K$_x$Fe$_2$As$_2$. \cite{KFe2As2_1,KFe2As2_2,KFe2As2_3} We also notice that the incipient $s_\pm$-wave is stabilized by the lurking top-flattened $\ga$ pocket, which enhances the virtual scattering between $\al$ and $\ga$ pockets down to moderate energy scales.

\begin{figure}[t]
\includegraphics[width=9cm]{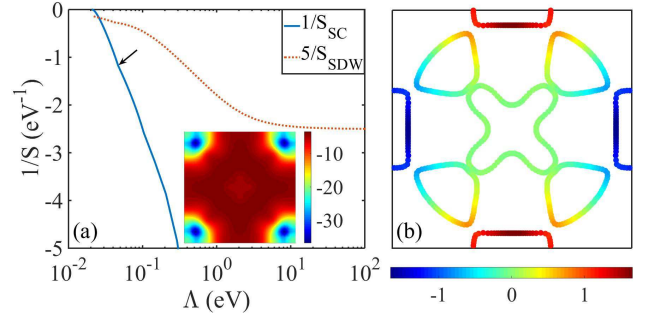}
\caption{(Color online) The same plot as Fig.\ref{ssc} but for $n=4.35$. The arrow
in (a) indicates the level crossing of the leading pairing channel.}\label{dsc}
\end{figure}

However, with further electron doping, the $\ga$ pocket sinks further below the Fermi level, and the effect of $\al$-$\ga$ scattering eventually becomes weaker than that between the $\bt$ pockets. As a result, $d_{x^2-y^2}$-wave pairing  on the $\bt$ pockets begin to dominate, as seen in Fig.\ref{dsc} for band filling $n =4.35$. We find that
the SDW channel is similar to the case for $n=4.24$, but in the SC channel, the interaction is strong in the incipient $s_\pm$-wave eigen mode at high energy scales, but the $d_{x^2-y^2}$-wave
becomes dominant at lower energy scales. The level crossing is indicated by the arrow in Fig.\ref{dsc}(a).
Fig.\ref{dsc}(b) shows the gap function in the band basis, with obvious $d_{x^2-y^2}$-wave symmetry. From the gap amplitude distribution, we see the active orbital for this pairing state is $d_{xy}$ since the gap function vanishes where the $d_{xy}$ weight of the normal state [as shown in Fig.\ref{fs}(d)] is small. More  orbital-resolved details of the pairing form factor can be found in the Appendix \ref{form}, which may be helpful in orbital-based mean field calculations.

\begin{figure}
\includegraphics[width=9.3cm]{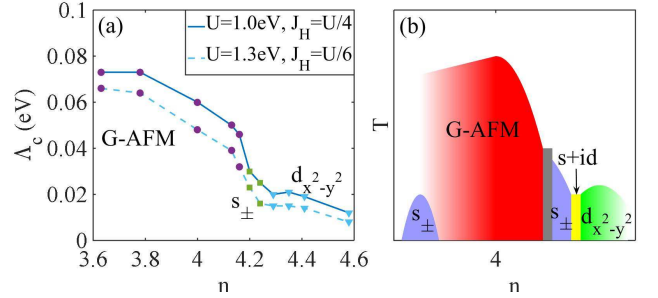}
\caption{(Color online)(a) The FRG diverging energy scale $\La_c$ plotted
as a function of band filling. The circles, squares, and triangles represent
$\La_c$ associated with the G-type AFM, incipient $s_{\pm}$-wave pairing, and $d_{x^2-y^2}$-
wave paring states, respectively. $U=1$eV, $J_H=U/4$ and $U=1.3$eV, $J_H=U/6$ for
solid and dashed lines, respectively. (b) A schematic temperature-doping phase
diagram for doped LaOCrAs. The gray region denote the transition between G-type
AFM state and incipient $s_{\pm}$-wave SC state. The dome in the hole doped case indicates
the possible $s_{\pm}$-wave SC state.} \label{pd}
\end{figure}

We have performed systematic calculations for various band fillings
around $n=4$. Fig.\ref{pd}(a) shows the critical energy scale $\La_c$
as a function of band filling. We find that the parent compound LaOCrAs
has a G-type AFM order, which is consistent with the powder neutron diffraction. \cite{LnOCrAs}
With the increase of electron-doping, the G-type AFM state is suppressed,
and the system develops incipient $s_\pm$-wave SC as the band filling is above $4.2$
electrons per site. Upon further electron doping, the system enters the $d_{x^2-y^2}$-wave SC state.
We also checked the hole-doped case to find that the G-type AFM states
is more robust (under the same interactions). This is due to the enhancement from
scattering between $\al$ and $\del$ pockets, and between the flat bands
near Fermi level around $\Ga$ and $M$ points.
The above results are not changed qualitatively for $ J_H \in [1/6, 1/4]U$
and moderate $U$. As a typical example, we set $U=1.3$eV, $J_H=U/6$, and
perform the FRG calculations. We find that the results are qualitatively
the same as the cases with $U=1$eV, $J_H=U/4$.

We end by presenting a schematic phase diagram for LaOCrAs in the
temperature-doping plane in Fig.\ref{pd}(b). There are two domes of
superconductivity in electron doped LaOCrAs. With the increase of
electron doping, the G-type AFM state gives way to the incipient $s_\pm$-wave SC state,
and subsequently to the $d_{x^2-y^2}$-wave SC state. This is somehow similar to the
two domes of superconductivity in electron-doped LaOFeAs with $H^-$
substitution of $O^{2-}$.\cite{LaOHFeAs} In the intervening regime between the
two domes, the two SC states may become nearly degenerate, and we anticipate a
time-reversal-symmetry breaking $s+id$-wave SC state is energetically favorable,
since the gap function would be maximally open on all fermi pockets.
To check this, we perform mean-field calculations to find that
$s+id$-wave SC state is indeed the ground state between the two domes.
(See Appendix \ref{mf} for details.) Finally,
in the hole-doped side, we find the $s_{\pm}$-wave SC may
replace the AFM state if the bare interaction is reduced to $U=0.7$eV, $J_H = U/4$
(or $U=0.9$eV, $J_H=U/6$). Following the argument that the effective bare interaction
may be reduced significantly far from the the Mott limit with $n=5$,\cite{LaOCrAs,Ishida}
we propose that the hole doped LaOCrAs may support $s_{\pm}$-wave superconductivity.

\section{Summary}\label{summary}

We constructed an effective five-band model for Cr $3d$ orbitals, and
investigated possible correlation-driven superconductivity in doped LaOCrAs.
With increasing electron doping, we find G-type AFM, incipient $s_\pm$-wave SC and
$d_{x^2-y^2}$-wave SC states. The gap function is highly anisotropic on the Fermi surface
and highly orbital-selective. Inbetween the two SC phases a time-reversal-symmetry breaking
$s+id$ SC phase is likely to occur. We also propose that the hole-doped LaOCrAs may support
the $s$-wave SC state.

We notice that superconductivity has not yet been observed experimentally
after electron doping in LaOCrAs through $F^-$ substitution of $O^{2-}$ with
band filling up to $n=4.2$. \cite{LnOCrAs} This is however consistent with our results, since the
superconductivity phase appears at $n > 4.2$ in our phase diagram. Therefore we propose further
electron doping in experiment. For this purpose, substitution of $H^-$ in place of $O^{2-}$ deserves
attention. This type of doping succeeds in iron oxypnictides in covering a very wide
doping range containing two SC domes. \cite{LaOHFeAs, CeOHFeAs, SmOHFeAsP}
Pressure may further help bring about superconductivity near the band filling $4.2$.

\acknowledgments{The project was supported by NSFC (under grant Nos.11604168,
11574134, 11404383 and 11604303) and the Ministry of Science and Technology of China (under grant No. 2016YFA0300401).
WSW also acknowledges the supports by Zhejiang Open Foundation of the Most Important Subjects (under grant No. xkzwl1613)
and K. C. Wong Magna Fund in Ningbo University.}\\

\appendix
\section{Form factors and gap function in the band basis} \label{form}
In this appendix we give the expressions of the form factors, or pairing matrix, $\phi_{SC}(\v k)$
in the orbital basis. To describe the momentum dependence, we introduce the lattice harmonics
\eqa c_x &=& \cos \v k_x ,  c_y = \cos \v k_y ; \nn
     s_x &=& \sin \v k_x ,  s_y = \sin \v k_y .
\eea
The non-vanishing elements of $\phi_{SC}(\v k)$ for $s_{\pm}$-wave pairing with band
filling $n=4.24$ are given by:
\eqa
  \phi^{11}_{SC}(\v k) &=& 0.13 + 0.95(c_x+c_y) + 0.44c_x c_y , \nn
  \phi^{22}_{SC}(\v k) &=& -0.05 + 0.06c_x + 0.05c_y - 0.02c_x c_y , \nn
  \phi^{33}_{SC}(\v k) &=& -0.05 + 0.05c_x + 0.06c_y - 0.02c_x c_y , \nn
  \phi^{44}_{SC}(\v k) &=& -0.07 + 0.02(c_x+c_y) - 0.01c_x c_y , \nn
  \phi^{55}_{SC}(\v k) &=& -0.03 + 0.03(c_x+c_y) - 0.04c_x c_y , \nn
  \phi^{12}_{SC}(\v k) &=& -\phi^{21}_{SC}(\v k) = 0.05is_x + 0.11is_x c_y , \nn
  \phi^{13}_{SC}(\v k) &=& -\phi^{31}_{SC}(\v k) = 0.05is_y + 0.11ic_x s_y , \nn
  \phi^{14}_{SC}(\v k) &=& \phi^{41}_{SC}(\v k) = -0.02(c_x - c_y) , \nn
  \phi^{23}_{SC}(\v k) &=& \phi^{32}_{SC}(\v k) = 0.01 c_x c_y , \nn
  \phi^{24}_{SC}(\v k) &=& -\phi^{42}_{SC}(\v k) = -0.01is_x , \nn
  \phi^{34}_{SC}(\v k) &=& -\phi^{43}_{SC}(\v k) = 0.01is_y , \nn
  \phi^{25}_{SC}(\v k) &=& -\phi^{52}_{SC}(\v k) = 0.01is_y - 0.03ic_x s_y, \nn
  \phi^{35}_{SC}(\v k) &=& -\phi^{53}_{SC}(\v k) = 0.01is_x - 0.03is_x c_y.
\eea
The non-vanishing elements of $\phi_{SC}(\v k)$ for $d_{x^2-y^2}$-wave
pairing with band filling $n=4.35$ are given by:
\eqa
  \phi^{11}_{SC}(\v k) &=& 0.02(c_x - c_y) , \nn
  \phi^{22}_{SC}(\v k) &=&  0.17 + 0.10c_x - 0.01c_x c_y , \nn
  \phi^{33}_{SC}(\v k) &=& -0.17 - 0.10c_y + 0.01c_x c_y , \nn
  \phi^{44}_{SC}(\v k) &=& 0.03(c_x - c_y) , \nn
  \phi^{55}_{SC}(\v k) &=& 0.85(c_x - c_y) , \nn
  \phi^{12}_{SC}(\v k) &=& -\phi^{21}_{SC}(\v k) = -0.04 i s_x + 0.02 i s_x c_y , \nn
  \phi^{13}_{SC}(\v k) &=& -\phi^{31}_{SC}(\v k) = 0.04 i s_y - 0.02 i c_x s_y , \nn
  \phi^{14}_{SC}(\v k) &=& \phi^{41}_{SC}(\v k) = 0.07 + 0.01(c_x+c_y)-0.01c_x c_y , \nn
  \phi^{24}_{SC}(\v k) &=& -\phi^{42}_{SC}(\v k) = 0.01 i s_x , \nn
  \phi^{34}_{SC}(\v k) &=& -\phi^{43}_{SC}(\v k) = 0.01 i s_y , \nn
  \phi^{25}_{SC}(\v k) &=& -\phi^{52}_{SC}(\v k) = -0.28 i s_y - 0.17 i c_x s_y, \nn
  \phi^{35}_{SC}(\v k) &=& -\phi^{53}_{SC}(\v k) = 0.28 i s_x + 0.17 i s_x c_y, \nn
  \phi^{45}_{SC}(\v k) &=& \phi^{54}_{SC}(\v k) = -0.09 s_x s_y.
\eea
With the pairing matrix $\phi_{SC}$ in the orbital basis, the gap function in the band basis is given by
$\Del_{\v kn} = \<\v kn|\phi_{SC}(\v k)|\v kn\>$ where $n$ is the band label and $|\v kn\>$ is the Bloch state in the given band. Here we used the fact that the normal-state hamiltonian is time-reversal-invariant.

\section{Mean field calculations in the superconducting phase} \label{mf}

\begin{figure}
	\includegraphics[width=8.5cm]{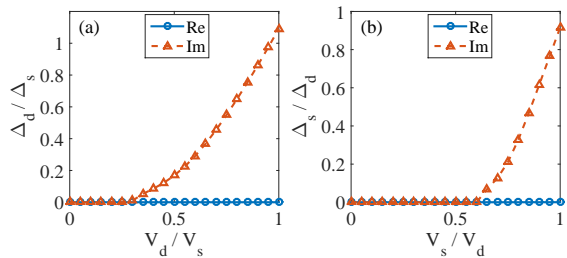}
	\caption{(Color online)(a) $\Del_d / \Del_s$ versus $V_d / V_s$
		with fixed $V_s = -0.3$eV. The solid and dashed lines are for the real and image parts of $\Del_d / \Del_s$.
		(b) $\Del_s/\Del_d$ versus $V_s/V_d$ with fixed $V_d = -0.3$eV.
	} \label{sid}
\end{figure}

If both the $s_{\pm}$-wave and $d_{x^2-y^2}$-wave states are dominant, the effective
low-energy Hamiltonian is given by
\eqa
  H = H_0 + \frac{V_s}{N}\sum_{\v k, \v k'} B^\dag_{s,\v k} B_{s,\v k'}
  +\frac{V_d}{N}\sum_{\v k, \v k'} B^\dag_{d,\v k} B_{d,\v k'}, \eea
where $H_0$ is the normal state dispersion, and $V_{s/d} < 0$ are the pairing interactions for $s_{\pm}$-wave
and $d_{x^2-y^2}$-wave pairing, respectively. $N$ is the number of
lattice sites, and $B^\dag_{s/d,\v k}$ are the pairing operator
\eqa B^\dag_{s/d, \v k} = \Psi^\dag_{\v k\ua} \phi^{s/d}_{SC}(\v k)(\Psi^\dag_{-\v k\da})^T , \eea
where $\Psi_{\v k\si}^\dag$ is a spinor creation field for all orbital degrees of freedom and $\si=\ua,\da$.
Since the form factor of a given symmetry changes slowly versus the doping level, we
use the form factors $\phi^{s/d}_{SC}(\v k)$ given in Appendix \ref{form} for simplicity.
The mean-field Hamiltonian can be written as
\eqa H_{MF} = H_0 + \sum_{\v k} [(\Del_s B^\dag_{s,\v k} + \Del_d B^\dag_{d, \v k}) + {\rm h.c.}], \eea
subject to the self-consistent conditions
\eqa \Del_{s/d} = \frac{V_{s/d}}{N}\sum_{\v k} \< B_{s/d,\v k} \>. \eea
In the calculations at zero temperature, out of $V_{s}$ and $V_d$, we fix one of them so that the mean field $T_c$ is
of the same order of the FRG divergence scale in the corresponding phase,
and take the other as a parameter for illustration.
We present $\Del_d / \Del_s$ versus $V_d / V_s$
with fixed $V_s = -0.3$eV in Fig.\ref{sid}(a), and $\Del_s / \Del_d$ versus $V_s / V_d$
with fixed $V_d = -0.3$eV in Fig.\ref{sid}(b). It is clear that around
$V_s=V_d$ the system develops time-reversal breaking $s+id$-wave SC state. Notice
that because of different rates of flow, a comparable $V_s$ and $V_d$ occurs in FRG only between the two superconducting domes discussed in the main text.


\begin{references}


\bibitem{Hosono1}
Y. Kamihara, T. Watanabe, M. Hirano, and H. Hosono, J. Am. Chem. Soc. {\bf 130}, 3296  (2008).

\bibitem{Ni1} 
T. Watanabe, H. Yanagi, T. Kamiya, Y. Kamihara, H. Hiramatsu, M. Hirano, and H. Hosono,
Inorg. Chem. {\bf 46}, 7719 (2007).

\bibitem{Ni2} 
T. Watanabe, H. Yanagi, Y. Kamihara, T. Kamiya, M. Hirano, and H. Hosono,
J. Solid State Chem. {\bf 181}, 2117 (2008).

\bibitem{Ni3} 
E. D. Bauer, F. Ronning, B. L. Scott, and J. D. Thompson, Phys. Rev. B {\bf 78}, 172504 (2008).

\bibitem{Ni4} 
F. Ronning, N. Kurita, E. D. Bauer, B. L. Scott, T. Park, T. Klimczuk, R. Movshovich,
and J. D. Thompson, J. Phys.-Condens. Mat. {\bf 20}, 342203 (2008).

\bibitem{Ni5} 
F. Ronning, E. D. Bauer, T. Park, S.-H. Baek, H. Sakai, and J. D. Thompson,
Phys. Rev. B {\bf 79}, 134507 (2009)

\bibitem{Ni6} 
Y. Tomioka, S. Ishida, M. Nakajima, T. Ito, H. Kito, A. Iyo, H. Eisaki, and S. Uchida,
Phys. Rev. B {\bf 79}, 132506 (2009).

\bibitem{Rh1} 
D. Hirai, T. Takayama, R. Higashinaka, H. A. Katori, and H. Takagi,
J. Phys. Soc. Jpn. {\bf 78}, 023706 (2009).

\bibitem{Rh2} 
N. Berry, C. Capan, G. Seyfarth, A. D. Bianchi, J. Ziller, and Z. Fisk,
Phys. Rev. B {\bf 79}, 180502(R) (2009).

\bibitem{Ir} 
D. Hirai, T. Takayama, D. Hashizume, R. Higashinaka, A. Yamamoto, A. Hiroko,
and H. Takagi, Physica C: Supercond. {\bf 470}, S296 (2010).


\bibitem{Pd1} 
V. K. Anand, H. Kim, M. A. Tanatar, R. Prozorov, and D. C. Johnston,
Phys. Rev. B {\bf 87}, 224510 (2013).

\bibitem{Pd2} 
Q. Guo, J. Yu, B. B. Ruan, D. Y. Chen, X. C. Wang, Q. G. Mu, B. J. Pan,
G. F. Chen, Z. A. Ren, EuroPhys. Lett. {\bf 113}, 17002 (2016).

\bibitem{Pt1} 
K. Kudo, Y. Nishikubo, and M. Nohara, J. Phys. Soc. Jpn. {\bf 79}, 123710 (2010).

\bibitem{Pt2} 
M. Imai, S. Emura, M. Nishio, Y. Matsushita, S. Ibuka, N. Eguchi, F. Ishikawa,
Y. Yamada, T. Muranaka, and J. Akimitsu, Supercond. Sci. Technol. {\bf 26},
075001 (2013).



\bibitem{Mn1}  
Y. Singh, A. Ellern, and D. C. Johnston, Phys. Rev. B {\bf 79}, 094519 (2009).

\bibitem{Mn2} 
Y. Singh, M. A. Green, Q. Huang, A. Kreyssig, R. J. McQueeney, D. C. Johnston, and A. I. Goldman,
Phys. Rev. B {\bf 80}, 100403 (2009)


\bibitem{Mn3} 
R. Nath, V. O. Garlea, A. I. Goldman, and D. C. Johnston, Phys. Rev. B {\bf 81}, 224513 (2010).

\bibitem{Mn4} 
N. Emery, E. J. Wildman, J. M. S. Skakle, A. C. Mclaughlin, R. I. Smith, and
A. N. Fitch, Phys. Rev. B {\bf 83}, 144429 (2011).


\bibitem{Mn5} 
A. T. Satya, A. Mani, A. Arulraj, N. V. Chandra Shekar, K. Vinod, C. S. Sundar, and
A. Bharathi, Phys. Rev. B {\bf 84}, 180515 (2011).


\bibitem{Mn6} 
Abhishek Pandey, R. S. Dhaka, J. Lamsal, Y. Lee, V. K. Anand, A. Kreyssig, T. W. Heitmann,
R. J. McQueeney, A. I. Goldman, B. N. Harmon, A. Kaminski, and D. C. Johnston
Phys. Rev. Lett. {\bf 108}, 087005 (2012).


\bibitem{Mn7} 
J. Lamsal, G. S. Tucker, T. W. Heitmann, A. Kreyssig, A. Jesche, A. Pandey, W. Tian,
R. J. McQueeney, D. C. Johnston, and A. I. Goldman, Phys. Rev. B {\bf 87}, 144418 (2013).


\bibitem{LaOMnP}
J. Guo, J. Simonson, L. Sun, Q. Wu, P. Gao, C. Zhang, D. Gu, G. Kotliar,
M. Aronson, and Z. Zhao, Sci. Reports \v 3, 2555 (2013).


\bibitem{LaOHMnAs}
T. Hanna, S. Matsuishi, K. Kodama, T. Otomo, S. I. Shamoto, and H. Hosono,
Phys. Rev. B {\bf 87}, 020401(R) (2013).


\bibitem{LaOCrAs}
J. M. Pizarro, M. J. Calder\'{o}n, J. Liu, M. C. Mu\~{n}oz, and
E. Bascones, arXiv: 1610.09560.

\bibitem{BaCr2As2_3}
M. Edelmann, G.Sangiovanni, M. Capone, and L. de' Medici, arXiv: 1610.10054.

\bibitem{CrAs1}
W. Wu, J. Cheng, K. Matsubayashi, P. Kong, F. Lin, C. Jin, N. Wang, Y. Uwatoko,
and J. Luo, Nat. Commun. {\bf 5}, 5508 (2014).

\bibitem{CrAs2}
H. Kotegawa, S. Nakahara, H. Tou, and H. Sugawara, J. Phys. Soc. Jpn. {\bf 83},093702 (2014).

\bibitem{ACrAs1}
J. K. Bao, J. Y. Liu, C. W. Ma, Z. H. Meng, Z. T. Tang, Y. L. Sun, H. F. Zhai,
H. Jiang, H. Bai, C. M. Feng, Z. A. Xu, and G. H. Cao, Phys. Rev. X {\bf 5}, 011013 (2015).

\bibitem{ACrAs2}
Z. T. Tang, J. K. Bao, Y. Liu, Y. L. Sun, A. Ablimit, H. F. Zhai, H. Jiang, C. M. Feng,
Z. A. Xu, and G. H. Cao, Phys. Rev. B {\bf 91}, 020506 (2015).

\bibitem{ACrAs3}
Z. T. Tang, J. K. Bao, Z. Wang, H. Bai, H. Jiang, Y. Liu, H. F. Zhai, C. M. Feng,
Z. A. Xu, and G. H. Cao, Sci. Chin. Mater. {\bf 58}, 16 (2015).


\bibitem{BaCr2As2_1}
M. Pfisterer and G. Nagorsen, Z. Naturforsch. B {\bf 35}, 703 (1980).

\bibitem{BaCr2As2_2}
D. J. Singh, A. S. Sefat, M. A. McGuire, B. C. Sales, D. Mandrus,
L. H. VanBebber, and V. Keppens, Phys. Rev. B {\bf 79}, 094429 (2009).

\bibitem{BaFe2As2}
M. Rotter, M. Tegel, and D. Johrendt, Phys. Rev. Lett. {\bf 101} 107006 (2008).


\bibitem{EuCr2As2}
U. B. Paramanik, R. Prasad, C. Geibel, and Z. Hossain, Phys. Rev. B {\bf 89},
144423 (2014).

\bibitem{LnOCrAs}
S. W. Park, H. Mizoguchi, K. Kodama, S. i. Shamoto, T. Otomo, S. Matsuichi,
T. Kamiya, and H. Hosono, Inorg. Chem. {\bf 52}, 13363 (2013).

\bibitem{SrCrAsO}
H. Jiang, J. K. Bao, H. F. Zhai, Z. T. Tang, Y. L. Sun, Y. Liu, Z. C. Wang,
H. Bai, Z. A. Xu, and G. H. Cao, Phys. Rev. B {\bf 92}, 205107 (2015).



\bibitem{wws1}
W.-S. Wang, Y.-Y. Xiang, Q.-H. Wang, F. Wang, F. Yang, and D.-H.
Lee, Phys. Rev. B {\bf 85}, 035414 (2012).

\bibitem{xyy1}
Y.-Y. Xiang, W.-S. Wang, Q.-H. Wang, and D.-H. Lee, Phys. Rev. B
{\bf 86}, 024523 (2012).

\bibitem{xyy2} Y.-Y. Xiang, F. Wang, D. Wang, Q.-H. Wang, and D.-H. Lee,
Phys. Rev. B {\bf 86}, 134508 (2012).

\bibitem{wws2}
W.-S. Wang, Z.-Z. Li, Y.-Y. Xiang, and Q.-H. Wang, Phys. Rev. B {\bf 87} 115135 (2013).

\bibitem{xyy3}
Y.-Y. Xiang, Y. Yang, W.-S. Wang, Z.-Z. Li, Q.-H. Wang, Phys. Rev. B {\bf 88}, 104516 (2013).

\bibitem{yy}
Y. Yang, W.-S. Wang, Y.-Y. Xiang, Z.-Z. Li, Q.-H. Wang, Phys. Rev. B {\bf 88}, 094519 (2013).

\bibitem{sro}
Q. H. Wang, C. Platt, Y. Yang, C. Honerkamp, F. C. Zhang, W. Hanke, T. M. Rice, R. Thomale, Europhys. Lett. {\bf 104}, 17013 (2013).

\bibitem{yy2}
Y. Yang, W.-S. Wang, J.-G. Liu, H. Chen, J.-H. Dai and Q.-H. Wang, Phys. Rev. B {\bf 89}, 094518 (2014).

\bibitem{wws3}
W.-S. Wang, Y. Yang, Q. H. Wang, Phys. Rev. B {\bf 90}, 094514 (2014).


\bibitem{LDA}
P. Giannozzi, \emph{et al.}, J.Phys.:Condens.Matter, {\bf 21}, 395502 (2009);
http:// www.quantum- espresso.org; Here we adopt the exchange correlation functional
introduced by J. P. Perdew, K. Burke and M. Ernzerhof [Phys. Rev. Lett.{\bf 77},  3865 (1996)],
and the wave functions are expanded by plane waves up to a cutoff energy of 45 Ry.
Also, $20 \times 20 \times 10$ $\v k$-point meshes are used with the special points
technique by H. J. Monkhorst and J. D. Pack [Phys. Rev. B {\bf 13}, 5188 (1976)].
The high-throughput ultrasoft pseudopotentials given in GBRV database are used
in our first-principles calculation [K. F. Garrity, J. W. Bennett, K. M. Rabe
and D. Vanderbilt, Comput. Mater. Sci. {\bf 81}, 446 (2014)]

\bibitem{Wannier}
N. Marzari and D. Vanderbilt, Phys. Rev. B {\bf 56}, 12847 (1997);
I. Souza, N. Marzari, and D. Vanderbilt, ibid. {\bf 65},
035109 (2001); A. A. Mostofi, J. R. Yates, G. Pizzi, Y. S. Lee,
I. Souza, D. Vanderbilt, and N. Marzari, Comput. Phys. Commun. {\bf 185},
2309(2014); The Wannier functions are generated by the
code developed by A. A. Mostofi, J. R. Yates, N. Marzari,
I. Souza, and D. Vanderbilt, http://www.wannier.org/.

\bibitem{Kuroki}
The main scheme is the same as K. Kuroki, S. Onari, R. Arita, H. Usui, Y. Tanaka,
H. Kontani, and H. Aoki, Phys. Rev. Lett. {\bf 101}, 087004 (2008).

\bibitem{wanyuan}
Notice that upon point group operations the orbitals also change.
See, e.g., Y. Wan and Q. H. Wang, Europhys. Lett.{\bf 85}, 57007 (2009).



\bibitem{KFe2As2_1}
K. Okazaki, Y. Ota, Y. Kotani, W. Malaeb, Y. Ishida, T. Shimojima, T. Kiss,
S. Watanabe, C.-T. Chen, K. Kihou, C. H. Lee, A. Iyo, H. Eisaki, T. Saito, H. Fukazawa,
Y. Kohori, K. Hashimoto, T. Shibauchi, Y. Matsuda, H. Ikeda, H. Miyahara,
R. Arita, A. Chainani, and S. Shin, Science {\bf 337}, 1314 (2012).


\bibitem{KFe2As2_2}

Y. Ota, K. Okazaki, Y. Kotani, T. Shimojima, W. Malaeb, S. Watanabe, C.-T. Chen,
K. Kihou, C. H. Lee, A. Iyo, H. Eisaki, T. Saito, H. Fukazawa, Y. Kohori, and S. Shin,
Phys. Rev. B {\bf 89}, 081103(R) (2014).

\bibitem{KFe2As2_3}

N. Xu, P. Richard, X. Shi, A. van Roekeghem, T. Qian, E. Razzoli, E. Rienks,
G.-F. Chen, E. Ieki, K. Nakayama, T. Sato, T. Takahashi, M. Shi, and H. Ding,
Phys. Rev. B {\bf 88}, 220508(R) (2013).

\bibitem{LaOHFeAs}
S. Iimura, S. Matsuichi, H. Sato, T. Hanna, Y. Muraba, S. W. Kim, J. E. Kim,
M. Takata, and H. Hosono, Nat. Commun. \v 3, 943 (2012).

\bibitem{Ishida}
H. Ishida and A. Liebsch, Phys. Rev. B {\bf 81}, 054513 (2010).

\bibitem{CeOHFeAs}
S. Matsuishi, T. Hanna, Y. Muraba, S. W. Kim, J. E. Kim, M. Takata, S. I. Shamoto,
R. I. Smith, and H. Hosono, Phys. Rev. B {\bf 85}, 014514 (2012).

\bibitem{SmOHFeAsP}
S. Matsuishi, T. Maruyama, S. Iimura, H. Hosono, Phys. Rev. B {\bf 89}, 094510 (2014).

\end{references}
\end{document}